# COOLHUNTING FOR THE WORLD'S THOUGHT LEADERS

Karin Frick, Detlef Guertler,
GDI Gottlieb Duttweiler Institute
Langhaldenstrasse 21
CH-8803 Rüschlikon/Zurich
Switzerland
{karin.frick, detlef.guertler}@gdi.ch

Peter A. Gloor
MIT CCI
5 Cambridge Center
Cambridge MA 02138
pgloor@mit.edu

## ABSTRACT

Which thinkers are we guided by? A novel "Thought Leader Map" shows the select group of people with real influence who are setting the trends in the market for ideas. The influencers in philosophy, sociology, economics, and the "hard sciences" have been identified by a Delphi process, asking 50 thought leaders to name their peers. The importance of the influencers is calculated by constructing a co-occurrence network in the Blogosphere. Our main insight is that the era of the great authorities seems to be over. Major thought leaders are rare – the picture is composed of many specialists.[1]

## MOTIVATION

Who are the thought leaders shaping current discourse on the future of business and society? What are the new global perspectives and theories helping to drive social change and innovation?

Every year, numerous lists are published about the world's largest companies, the most promising start-ups, the strongest consumer brands, the richest individuals, the most successful sports stars, the top chefs and the most important trends in technology. Unlike these lists of business or technology leaders and trends, the most important thought leaders and trends shaping our society have not been subjected so far to any truly systematic analysis and regular publication – nothing remotely comparable to the analysis behind Gartner's technology trends, for example. And yet the market of "big" ideas also yields both innovators and trends, which guide the decisions made within politics and business, which influence public opinion and which inspire further research or attract investors – and which are therefore well worth monitoring. As a rule, the importance of individual thinkers is measured on the basis of frequency of citation (citation index), sales figures from non-fiction and reference works, and academic accolades (e.g. the Nobel Prize). Magazines such as Foreign Policy and Time publish annual lists of leading personalities from business, politics, research, art and culture, based on the results of polling experts for their opinions. And the TED.com website beautifully showcases the current leaders in the market for ideas, showing which talks with ideas for the future have had the most views and recommendations. While such rankings can give a rough idea of the popularity of individual thought leaders, they say little about the size of their actual influence, nor about the trendsetters in the market of ideas and how these ideas propagate. Nor do they show how the various thinkers and doctrines are interconnected – and who is influencing whom. New ideas are not created in a vacuum but in the act of engagement with a range of separate doctrines: accordingly, one must also consider these ideas in their juxtaposition to views held by other academics and researchers. A thinker gains influence only if his or her ideas attract attention, are taken on board by others and are then discussed in depth. This not only means discussions held within an inner circle or research group but also the wider, subsequent debate with a broader public, which also includes laypeople. Today, the most important marketplace for new ideas is the Internet, where they are first presented, disseminated and most vigorously debated. If we want to gauge the actual influence thinkers possess, we therefore need to assess their status on the net and the intensity of debate in the virtual infosphere about these people and their ideas.

## OUR APPROACH

To analyze the status and the popularity of selected thought leaders in the infosphere/blogosphere, we have been using the network analysis tool Condor (Gloor & Zhao 2004). The same method has been used for analyzing product brands, or measuring stocks, for example. We have been working with this approach since 2005 to compare the positioning of brands, companies, concepts or individuals in the infosphere and to produce graphics of this kind. The software itself doesn't care whether the subject is Pepsi vs. Coke, Obama vs. Romney or Krugman vs. Kahneman. We started our list by manually selecting a list of over 100 thinkers in the fields of philosophy, sociology, economics, and the "hard sciences" based on the assessment of three researchers based on extensive literature research. These thinkers were then ranked using the following process: To find the most significant pages about a certain thought leader involves the application of a subject-driven betweenness algorithm similar to Google's page rank. Condor collects the most important Blogs mentioning say "Daniel Kahnemann", and then plugs these URLs into a Web search engine to see which other blogs link back to them. While Google's page rank is fixed with each website having a predetermined value, our page rank is topic-based: for debates about ideas, the "Huffington Post" has very high relevance, for example – but very low relevance for conversations about dog food.

---

[1] This paper is an extended version of (Frick 2012) published in German in the GDI Impuls Magazine.

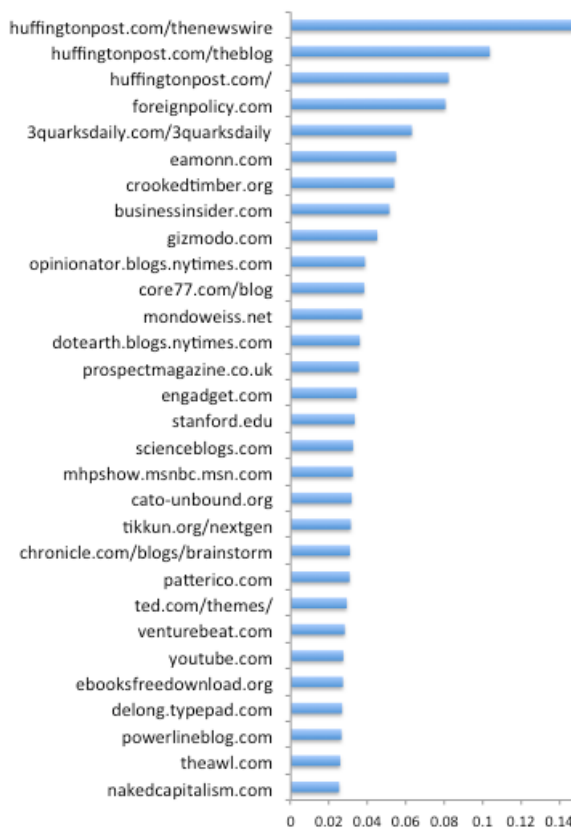

*Figure 1. Most important blogs for thought leaders (sorted by betweenness centrality)*

Figure 1 shows the most important Websites collected through this process, listing the Web sites from figure 2 by their betweenness in the network. While the search was run on the entire Web, the top scoring blogs were all English-language blogs. For the resulting visualization (figure 2) Condor is using a graph-layout algorithm based on force-directed spring layout, where two people are connected if they appear on the same Web page. To put it in other words: the more often two thinkers are named together in the infosphere, the closer together they are shown here in the graph. The size itself is determined by betweenness centrality, which is influenced by the citation frequency and by the relevance of the pages on which they are featured. Our Coolhunting tool Condor then generates this image by aggregating the links between the individuals analyzed.

Making use of the fact that – other than general Web sites – blogs are continuously updated, this method is very sensitive to the time when the blog queries have been conducted. When this analysis was carried out (August/September 2012), for example, Thilo Sarrazin had just published his most recent book "Europe doesn't need the euro" – and this had a positive effect on his ranking. Re-evaluating his position six months later, his position will certainly be quite different – and thinkers who are then being hotly debated will have a higher relevance. Comparisons over extended periods of time will let us distinguish between "one-hit thinkers" and long-term thought leaders.

**A FRAGMENTATION OF IDEAS**

The results of this network analysis present a highly fragmented picture. There are no thinkers who really dominate the landscape: the distance between the «stars» and the less significant and less well-known researchers is relatively small and presumably only temporary. Other analyses will reveal the degree to which the relative social network positions change over time. The era of the great authorities seems to be over. Instead of a handful of key thinkers, we see a broad spectrum of specialists, who focus on niche topics, who remain generally unknown outside their specialist field and whose work is not discussed (figure 2 and table 1 in the appendix). As with the market for books and films, the market for ideas also seems increasingly a niche market, where major ideas and their creators are now losing ground to minor ideas and unknown researchers. Attention is no longer focused on the next big idea or the next Einstein, but is now increasingly divided up among many small-scale ideas – the "long tail of ideas". This picture – namely the absence of authoritative thinkers and key concepts that influence whole generations of intellectuals – seemed so extraordinary that we tested it by means of a different assessment technique: we surveyed contributing authors to the journal GDI Impuls, a German knowledge and ideas magazine published by GDI. We asked these former authors for GDI Impuls (experts from both research and practice) to name the thinkers who have most strongly influenced their own work and also the persons who will be the thought leaders of the 21st century. Receiving no prior briefing, fifty experts from home and abroad (including such luminaries as were already ranked in our Thought Leader list) generated a list of around 300 different names. On this long list of the major thinkers of yesterday and tomorrow, multiple mentions are rare. Karl Marx (5), Niklas Luhmann (4) and Michel Foucault (3) are the most-cited thinkers from the past, whose ideas have most strongly influenced the work of our experts.

Daniel Kahneman (5), Anthony Giddens (3), Malcolm Gladwell (3) and Paul Krugman (3) top the list of the most important thinkers for the 21st century. All in all, then, this survey gives us the same picture as that from the network analysis: it's not about altitude, but latitude.

### Economists rule

A thought leader's importance depends on the one hand on whom you ask and on the other, on how one measures it. If we take the citation frequency in academic journals as our benchmark, we find the behavioral economist and Nobel Prize winner Daniel Kahneman also among the leaders (table 1 in the appendix), but the remaining positions in the Thought Leader map now look very different: Thilo Sarrazin, author and former Berlin Finance Senator, would no longer be ranked first, for example, but would be bringing up the rear. This seems to indicate that individuals who carry little weight in research circles may be ideas market trendsetters, however, and can – for a certain period of time – set the agenda of public debate. If we take Google search hits as our benchmark then our rankings once again change completely. Of the thinkers we consider, the physicist Stephan Hawking now has the most Google hits – although he is a mid-fielder in terms of network status. Only Daniel Kahneman achieves consistently high rankings. It appears, therefore, that a strong online presence does not automatically lead to the actual discussion of an idea – and thus to a greater influence on the zeitgeist – than ideas whose creators have fewer Google hits. The field of influential thinkers is broad and too large to present in its entirety on the Thought Leader map. If we consider our thought leaders' disciplines, however, and the universities at which they work, then the field contracts once again. The categories have been determined based on the thought leaders' Wikipedia pages and home pages. Discussions about the future are clearly dominated by economists (24), followed by political theorists (8), social theorists (7) and philosophers (5). Scientists, in contrast, are less represented in discussions about the future in purely quantitative terms. Yet these few individuals – biologists (5), computer scientists (4), physicists (3) and chemists

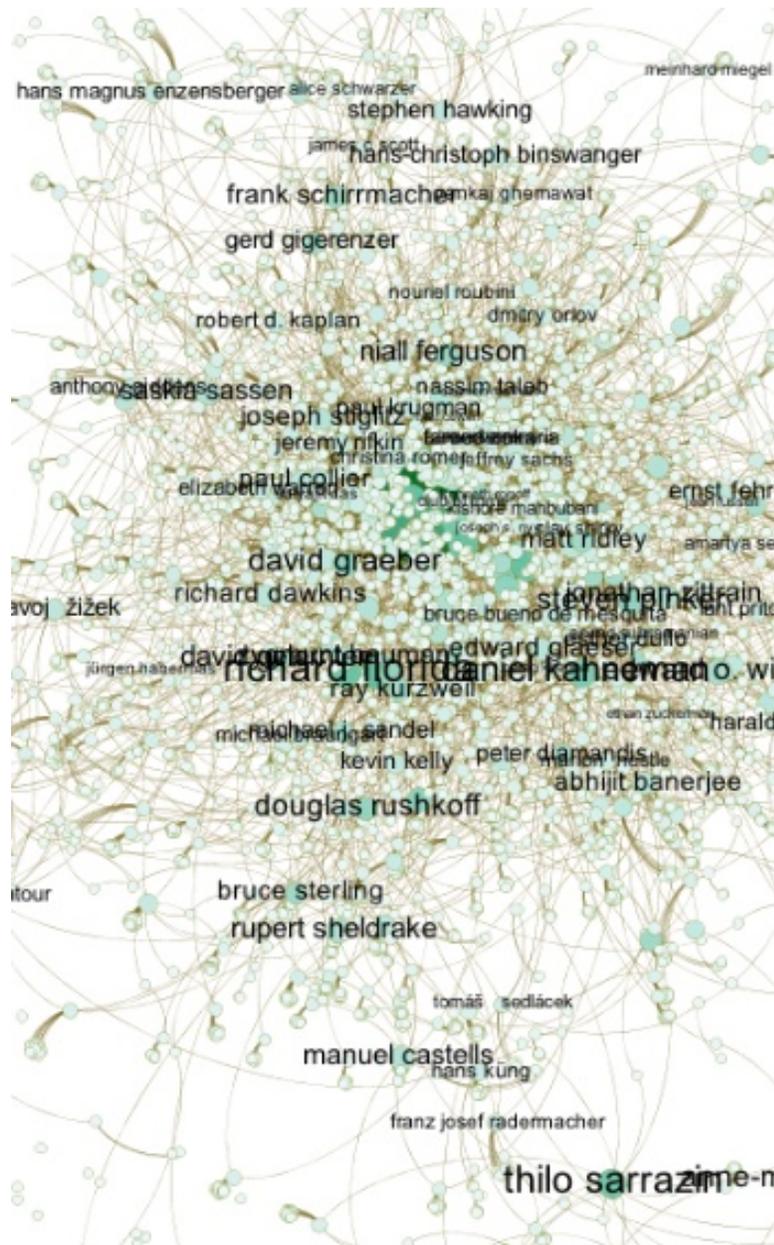

*Figure 2. Thought leader map*

(2) – are nonetheless all ranked in the upper half of the thought leader list, i.e. they enjoy above-average popularity in the blogosphere. Individual thinkers are not the only constituent parts of the network. Their environment – friends, research partners, students, sponsors and institutions where they work – also plays a part. Here, we note that the traditional elite universities continue to exert a powerful influence: Harvard (12) leads New York (6), Princeton, London (4 each), Yale and Columbia (3 each). While knowledge creation continues to become more and more open, and operates much like a bazaar – where many thinkers develop and exchange a wide variety of new ideas – the universities, as "cathedrals of learning", have clearly not lost their central importance (Raymond 1999).

today, still write a book that gets one noticed and triggers a debate that is clearly necessary for disseminating one's ideas. A book makes the idea tangible and durable, ensuring that people's engagement with the idea can develop and grow, both online and offline. The US social theorist Randall Collins (1998) has written what is probably the most comprehensive work on the formation of intellectual standpoints. One core aspect of his theory is that new ideas are always generated by the rivalry between contemporary thinkers, and that creativity is at its highest when there is an especially high level of friction between competing ideas. If one therefore assumes that new ideas are generated by engagement and debate, it is interesting to track the viewpoints between which these lines of conflict are drawn up today.

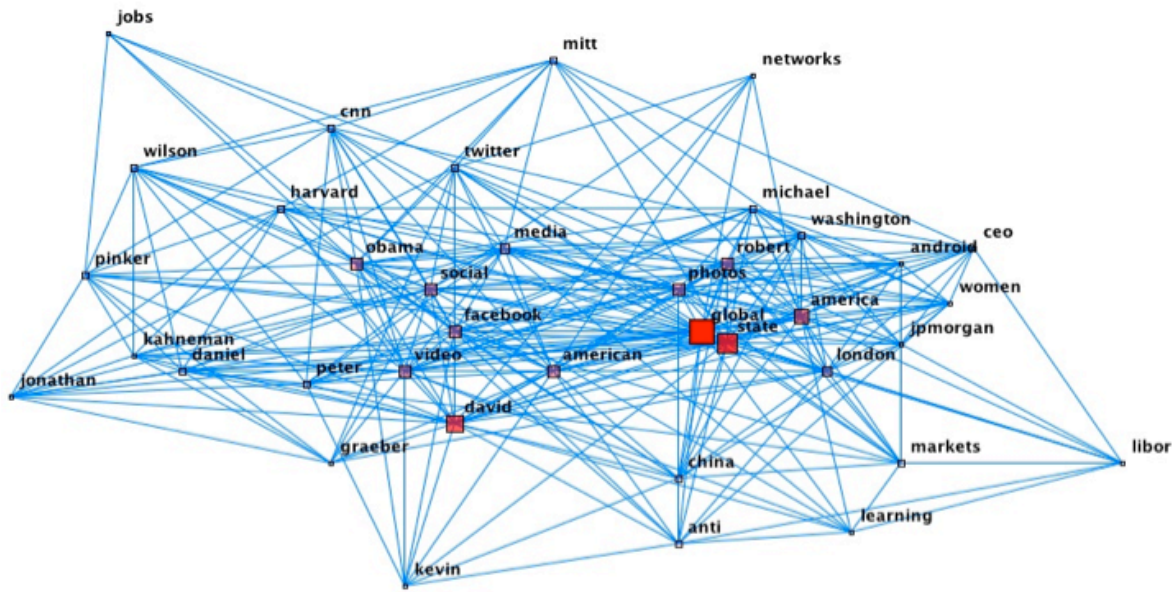

*Figure 3. Key terms in Blog posts about thinkers*

### Writing a Book is key

Figure 3 shows a tag cloud generated with Condor based on the text nearest to the names of the thinkers on the blog posts, illustrating the big role Harvard, Facebook, and the US President play in launching new ideas. At the same time we also notice a spillover effect from the banking crises, leading to a discussion of new social models and ethics. Thinkers who generate a strong response from the blogosphere and occupy a central position in the thought leader network are also successful authors in their own right and have produced one or more bestsellers in recent years. Out of the 76 thought leaders we surveyed, 74 have written at least one book. It seems, therefore, that the book continues to be the medium of choice for making one's ideas heard and achieving a central role in the knowledge market. Accordingly, it seems that anyone wishing to change the world must, even

### WIKIPEDIA ADDS A MULTICULTURAL DIMENSION

Our thinker short-list seems to favor the West. Viewed from India, the Middle East or China, the Thought Leader map might look quite different.
To compare key leaders across different cultures, we use a completely different set of sources. We therefore also applied another method (Kleeb et al. 2012) where we constructed a global link network using Wikipedia as the source. Articles about people in Wikipedia include many social cues about a particular person. In the English Wikipedia there is a category called "Living People" which includes articles of people currently alive. To construct the social network of living people, we collected each article's contents and extracted the internal links to other living people articles. Similarly to the Google page rank algorithm, this linking structure gives us a clue about which person is more prominent from the Wikipedians' perspective. As metric of importance we took the ratio between in-degree and out-degree because having a higher in-degree might be a proxy for the power or importance of the person described in the article. Because the size of the original map was too huge, we only included articles having more than a predefined number of incoming links and those peripheral articles that have a direct link to the most prominent articles.

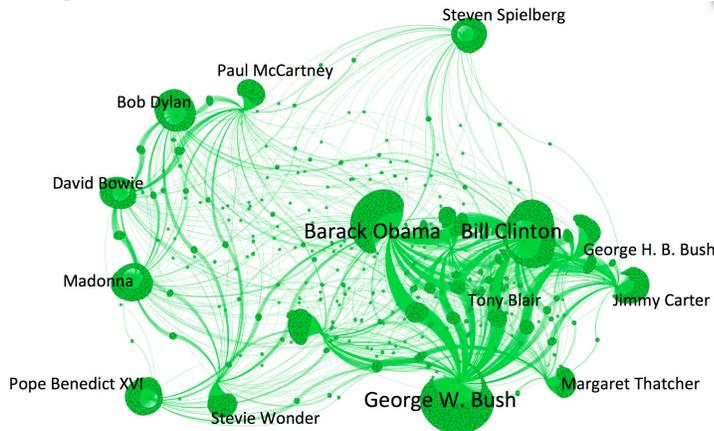

*Figure 4. World leaders according to the 2011 English Wikipedia (Kleeb 2011)*

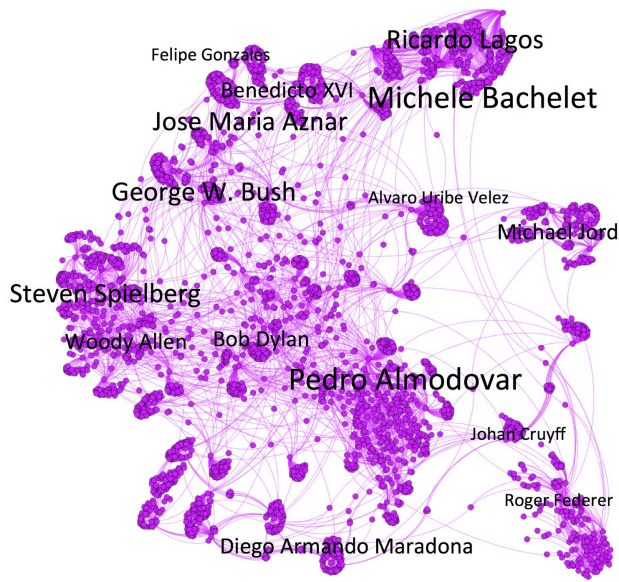

*Figure 5. World leaders according to the Spanish 2011 Wikipedia (Kleeb 2011)*

To draw a temporal map of the "living people network", we collected snapshots of the contents of the living people articles at different points in time. We also visualized living people networks in seven different-language Wikipedias (English, German, Spanish, French, Chinese, Arabic, Korean, Japanese) collecting the articles of people who were born after 1880 and not dead until 2010. Figure 4 shows the key leaders in the English Wikipedia, figure 5 the same for the Spanish Wikipedia.

Obviously the world view of Spanish speakers is rather different from native English speakers, with former Chilean president Michele Bachelet, and movie dirctor Pedro Almovodar occupying central positions.

## CONCLUSIONS

The new ideas show great diversity: the question of whether the field will continue to differentiate or consolidate can be answered only in the long term. The diversity of ideas and thinkers also reflects the complexity of the world itself, which is "too big to know" (Weinberger, 2012). Knowledge volume and rate of growth are now too great for the market to be dominated by just a few ideas. Never have there been so many researchers as today – nor has academic output ever been so diverse. We can no longer rely on the old measurement systems as a means of orienting ourselves and finding relevant, new ideas. Our Thought Leader map is an attempt to establish a new approach for classifying the most influential thinkers and trends in published research. We will continue to develop this ranking system and will conduct further network analyses at regular time intervals, combining our polling-based blog-ranking system with autogenerated maps from Wikipedia.

Experts Are Everywhere, and the Smartest Person in the Room Is the Room. Basic Books

| Name | Thought Leader Index | Google Citation H-Index | Google hits |
|---|---|---|---|
| Richard Florida | 0.15532562 | 58 | 815000 |
| Thilo Sarrazin | 0.034529135 | 7 | 3160000 |
| Daniel Kahneman | 0.032592207 | 131 | 3990000 |
| David Graeber | 0.025085267 | 36 | 1450000 |
| Steven Pinker | 0.024713721 | 64 | 1880000 |
| Douglas Rushkoff | 0.023806253 | 18 | 579000 |
| Niall Ferguson | 0.023014171 | 41 | 2380000 |
| David Gelernter | 0.022511605 | 39 | 293000 |
| Frank Schirrmacher | 0.02140022 | 19 | 499000 |
| Franz Josef Radermacher | 0.021215923 | 15 | 44100 |
| Ray Kurzweil | 0.020574821 | 14 | 1650000 |
| Bruce Sterling | 0.020151732 | 27 | 1160000 |
| Matt Ridley | 0.019829774 | 18 | 764000 |
| Gerd Gigerenzer | 0.019788496 | 69 | 230000 |
| Michael J. Sandel | 0.019454172 | 32 | 5500000 |
| Peter Diamandis | 0.017538087 | 12 | 462000 |
| Edward O. Wilson | 0.017428175 | 78 | 786000 |
| Anne-Marie Slaughter | 0.017274639 | 40 | 670000 |
| Rupert Sheldrake | 0.016360085 | 26 | 747000 |
| Manuel Castells | 0.015326205 | 79 | 1260000 |
| Saskia Sassen | 0.015142039 | 65 | 377000 |
| Zygmunt Bauman | 0.015125725 | 79 | 1490000 |
| Jonathan Zittrain | 0.01476404 | 19 | 389000 |
| James C. Scott | 0.013972453 | 91 | 1570000 |
| Edward Glaeser | 0.013923897 | 88 | 264000 |
| Joseph Stiglitz | 0.013834712 | undefined | 4490000 |
| Abhijit Banerjee | 0.013574838 | 58 | 285000 |
| Paul Collier | 0.013307666 | 78 | 1070000 |
| Richard Dawkins | 0.013153211 | 47 | 11300000 |
| Slavoj Žižek | 0.012257034 | 49 | 3470000 |
| Ernst Ulrich von Weizsäcker | 0.011952454 | 13 | 232000 |
| Ernst Fehr | 0.011791188 | 74 | 125000 |
| Kevin Kelly | 0.011517704 | undefined | 2410000 |
| Stephen Hawking | 0.011512456 | 74 | 13900000 |
| Hans-Christoph Binswanger | 0.011404731 | 14 | 102000 |
| Harald Welzer | 0.011213235 | 19 | 247000 |
| Jeremy Rifkin | 0.011129971 | 44 | 1280000 |
| Paul Krugman | 0.010957496 | 63 | 10200000 |
| Esther Duflo | 0.01090141 | 47 | 228000 |
| Robert D. Kaplan | 0.010281591 | 26 | 397000 |
| Nassim Taleb | 0.010244179 | 20 | 598000 |

*Table 1. (Frick 2012) Top 41 key thinkers (out of 76) ranked by our approach, compared against citation index and search hits*